\documentclass[]{eps}

\usepackage{graphicx}
\usepackage{times}
\usepackage{natbib}

\volumenumber{xx}
\publishyear{2012}
\frompage{\pageref{firstpage}}
\topage{\pageref{finalpage}}

\newcommand{\commentpcf}[1]{}

\def\glon{\ell}
\def\glat{$b$}
\def\kms{km s$^{-1}$}
\def\cmtwo{cm$^{-2}$}
\def\cc{cm$^{-3}$}
\def\Rgd{$R_\mathrm{g/d} $}
\def\micron{$\mu$m}
\def\lesssim{\frac{<}{\sim}}
\def\microG{$\mu$G}

\def\deeg{$^\circ$}

\def\glon{$\ell$}
\def\glat{$b$}

\def\OI{O$^{\rm o}$}
\def\MgII{Mg$^{\rm +}$}
\def\FeII{Fe$^{\rm +}$}
\def\SiII{Si$^{\rm +}$}

\def\NHI{$N$(H$^\circ$)}
\def\HeI{He$^\circ$}
\def\NeI{Ne$^{\rm o}$}

\def\NHI{$N$(H$^{\rm o}$)}

\newcommand\aj{{AJ}}%
\newcommand\araa{{ARA\&A}}%
\newcommand\apj{{ApJ}}%
\newcommand\apjl{{ApJ}}%
\newcommand\apjs{{ApJS}}%
\newcommand\aap{{A\&A}}%
\newcommand\mnras{{MNRAS}}%
\newcommand\ssr{{Space~Sci.~Rev.}}%
\newcommand\planss{{Planet.~Space~Sci.}}%
\shorttitle{Frisch \lowercase{\textit{and}} Slavin: Local Interstellar Dust}

\title{Interstellar Dust Close to the Sun}
\author{Priscilla C. Frisch$^1$ and Jonathan D. Slavin$^2$}
\affiliation{$^1$Dept. Astronomy and Astrophysics, University of Chicago\\
             $^2$Harvard-Smithsonian Center for Astrophysics}
\received{February 9, 2012}\revised{May 2, 2012}\accepted{May 2, 2013}

\abstract{ The low density interstellar medium (ISM) close to the Sun
and inside of the heliosphere provides a unique laboratory for
studying interstellar dust grains.  Grain characteristics in the
nearby ISM are obtained from observations of interstellar gas and dust
inside of the heliosphere and the interstellar gas towards nearby
stars.  Comparison between the gas composition and solar abundances
suggests that grains are dominated by olivines and possibly some form
of iron oxide.  Measurements of the interstellar Ne/O ratio by the
Interstellar Boundary Explorer spacecraft indicate that a high
fraction of interstellar oxygen in the ISM must be depleted onto dust
grains.  Local interstellar abundances are consistent with grain
destruction in $\sim 150$ \kms interstellar shocks, provided that the
carbonaceous component is hydrogenated amorphous carbon and carbon
abundances are correct.  Variations in relative abundances of
refractories in gas suggest variations in the history of grain
destruction in nearby ISM.  The large observed grains, $> 1$ \micron,
may indicate a nearby reservoir of denser ISM.  Theoretical
three-dimensional models of the interaction between interstellar dust
grains and the solar wind predict that plumes of $\sim 0.18$ \micron\
dust grains form around the heliosphere.  } \keywords{interstellar,
dust, heliosphere, abundances}

\begin{document}
\label{firstpage}
\maketitle
\copyrighttext{}

\section{Introduction }

The local interstellar medium (ISM) surrounding the Sun and inside of
the heliosphere offers a unique set of diagnostics for interstellar
dust grains, in a well-defined region of space.  In the local low
density ISM, interstellar dust grains (ISDGs) can be studied far from
the dense clouds where grain mantle accretion occurs.  Starlight
reddening data show that the ISM opacity becomes significant only
beyond $\sim 70 -100$ pc \cite{Fitzgerald:1968}.  Because of the low
densities of the Local Bubble, $n<0.005$ \cc, interstellar shocks
propagate relatively long distances before dissipating.  Local clouds
are partially ionized and have low column densities which lead to the
need to include ionization corrections to calculate the gas phase
abundances of refractory elements and thereby infer the composition of
the dust \cite{Frisch:2011araa}.  Variations in the relative
abundances of Fe, Mg, and Si in the local ISM indicate variations in
the processing history of the dust over parsec-sized spatial scales.
In the absence of infrared-bright dense clouds, the composition of
local dust is inferred from measurements of interstellar gas towards
nearby stars and inside of the heliosphere.  \emph{In situ}
measurements of neutral interstellar gas and dust grains provide a
benchmark for understanding the Local Interstellar Cloud (LIC)
surrounding the heliosphere \cite{Frischetal:2009ibex,Bochsleretal:2012isn}.  The mass
distribution is given by \emph{in situ} observations of interstellar
dust in the heliosphere, at least for the larger grains.  A unique
local diagnostic of the gas-to-dust mass ratio, \Rgd, is provided by
comparisons between \emph{in situ} dust data and the local ISM
properties inferred from observations of interstellar gas inside and
outside of the heliosphere.  The observed deficit of small grains is
attributed to the interaction of ISDGs with the heliosphere, while the
large-grain excess provides fundamental information about the origin
and destruction of grains in the ambient ISM.

\section{Characteristics of local interstellar medium} \label{sec:lism}

\begin{figure}[!th]
\begin{center}
    \makebox{\includegraphics[width=0.4\textwidth]{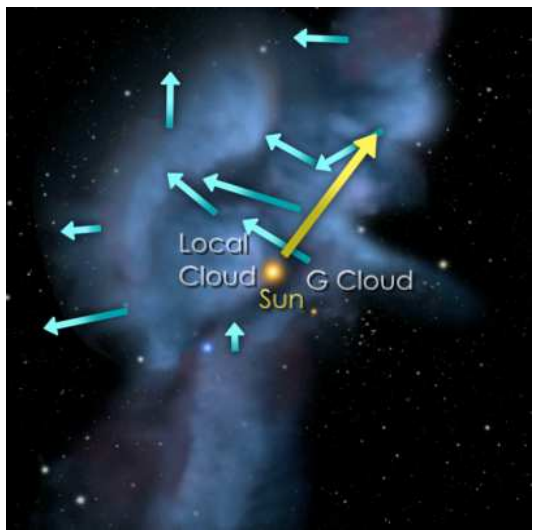}}
\end{center}
\vspace*{-5mm}
\caption[Cluster of local interstellar clouds:]{ Parts of the cluster
of local interstellar clouds within 10 pc are shown projected onto the
galactic plane.  The yellow dot shows the location of the Sun and the
yellow arrow gives the direction of solar motion through the LSR.
Cloud motions relative to the LSR are shown with blue arrows.  The
galactic center is towards the right of the figure, and \glon=90\deeg\
is at the figure top.  The bulk motion of these clouds is away from
the center of the Loop I superbubble.  For more information on the
cloud names and velocities see Fig. 11 in \cite{Frisch:2011araa} and
\cite{RLIV:2008vel}.  (Figure credit: Adler Planetarium, NASA,
Priscilla Frisch, Seth Redfield, Jeff Linsky) }
\label{fig:localfluff}
\end{figure}

The local ISM is clumped into low density clouds that flow past the
Sun with a mean bulk velocity of 17 \kms\ in the local standard of
rest (LSR), and away from the direction of the center of the Loop I
superbubble near \glon$ = 335$\deeg, \glat$= -5$\deeg.  The low local
densities, $<n_\mathrm{HI}> \sim 0.1$ \cc\ and \NHI$ < 10^{18.5}$
\cmtwo, correspond to extinctions of E(B-V) $<0.001 $ mag towards
nearby stars.  Interstellar dust within $\sim $15 pc of the Sun is the
topic of this discussion. Dust properties can be inferred from
observations of gas towards stars within $\sim 40$ pc, since most
nearby ISM within 40 pc is also within 15 parsecs, and also from
\emph{in situ} observations of interstellar dust inside the
heliosphere.  Local clouds are traced by optical and UV absorption
lines, and sorted into clouds based on common velocities (Figure
\ref{fig:localfluff}).  For a recent review of local ISM properties
see \cite{Frisch:2011araa}.

Over the past 15 Myrs, stellar winds and supernova from the nearby
stellar associations in Scorpius, Centaurus, and Orion have formed
superbubbles extending to the solar vicinity (e.g. Loop I and the
Orion superbubble).  Loop I is a giant ($\sim 90^\circ$ radius) radio
loop of polarized synchrotron emission located in the northern
hemisphere, and surrounding the Scorpius-Centaurus Association.  The
superbubble shell is about 4 Myrs old \cite{Frisch:1996}.  Models of
Loop I that assume it is a spherical object place the Sun in or near
the shell rim.  The bulk motion of the cluster of local interstellar
clouds (CLIC) is flowing from a position near the center of Loop I, as
defined by the S1 superbubble shell centered $78 \pm 10$ pc away at
\glon$=346^\circ \pm 5^\circ$, \glat$=-3^\circ \pm 5^\circ$
\cite{Frisch:2011araa,Wolleben:2007}.

Both the kinematics of the CLIC and the direction of the local
magnetic field suggest that the CLIC is associated with an evolved
superbubble shell.  The local field direction has been derived both
from the weak polarizations of nearby stars, 0.0001\%-0.02\%
\cite{Frisch:2010ismf1,Frisch:2011ismf2}, and from the interstellar
magnetic field direction defined by the ``ribbon'' of interstellar
neutral atoms observed by the Interstellar Boundary Explorer (IBEX)
spacecraft \cite{McComas:2009sci}.  These data indicate that the ISMF
within several parsecs is approximately parallel to the rim (or
perpendicular to a surface normal) of the S1 shell
\cite{Frisch:2010ismf1,Frisch:2011ismf2}.  Together these data suggest
that interstellar dust grains within several parsecs have recently
been subjected to interstellar shocks.

\section{Composition of local interstellar dust } \label{sec:composition}

In the low density local ISM, where dust emission is difficult to
observe, the interstellar dust composition is inferred from the
assumption that atoms missing from the gas-phase are depleted onto
dust grains.  This method requires knowledge of the chemical
composition of the ISM.  The isotopic composition of the LIC that
feeds ISM into the heliosphere indicates that the reference abundances
for the LIC may be solar.  LIC neutrals are ionized inside of the
heliosphere, forming pickup ions that are accelerated to become the
anomalous cosmic rays (energy $< 70$ MeV/nucleon).  Anomalous cosmic
rays have isotopic compositions that are approximately solar,
$\mathrm{^{20}Ne/^{22}Ne} \sim 13.7$ and $\mathrm{^{16}O/^{18}O} \sim
500$, instead of showing galactic cosmic ray ratios
\cite{Leske:2000isotope}.  X-ray spectroscopy of absorption edges in
X-ray binaries also show ratios of interstellar Ne/O that are
consistent with the solar value \cite{Juett:2006}.

Three means of determining the composition of local dust grains are
discussed, based on: (1) refractory elements that are missing from the
gas-phase; (2) photoionization models of the interstellar gas entering
the heliosphere, that reconstruct the depletion pattern of the gas;
(3) \emph{in situ} detections of interstellar Ne and O in the
heliosphere.  Methods 2 and 3 both indicate that a large fraction
(35\%--40\%) of the oxygen in the local ISM must be depleted onto dust
grains.

\subsection{Refractory elements in CLIC}

\begin{figure}[th]
\begin{center}
    \makebox{\includegraphics[width=0.5\textwidth]{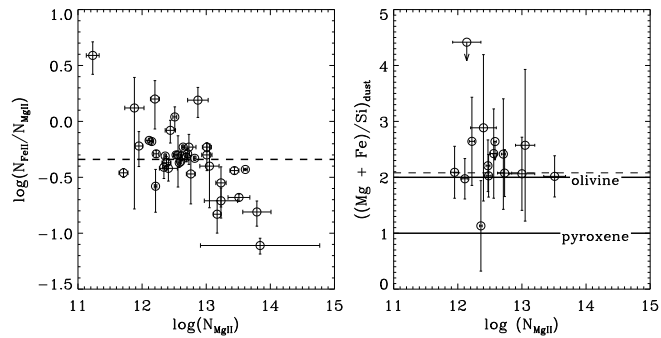}}

\end{center}
\vspace*{-5mm}
\caption[Refractory elements in local interstellar clouds:]{ The
observed gas-phase column densities of \MgII, \FeII, and \SiII\ in the
local ISM indicate that although there is a wide variation in the
relative amount of \FeII\ when compared to \MgII\ (left), the primary
dust component has (Mg+Fe)/Si$\sim 2$ such as expected for olivine
silicates (right).  The comparable value for pyroxene is
(Mg+Fe)/Si$\sim 1$.  Grain destruction returns more Mg than Fe to the
gas-phase.  The dashed line in the left figure shows the ratio
predicted by photoionization models of the LIC
\cite{SlavinFrisch:2008}, where it is also predicted that 15\% of the
Mg is doubly ionized \cite{SlavinFrisch:2008}.  The dust composition
on the right is obtained by comparing the gas-phase abundances of the
elements with solar abundances.  The presence of iron oxides in the
dust would help explain the highest ratios (This figure is courtesy of
Seth Redfield, using results from \cite{RLI,RLII,Frisch:2011araa}.).
}
\label{fig:mgfe}
\end{figure}

Surveys of interstellar absorption lines towards nearby stars that
sample the CLIC (e.g. \cite{RLI,RLII} and references therein) have
shown that the relative gas-phase abundances of Fe and Mg vary within
15 pc of the Sun (Figure \ref{fig:mgfe}, left).  A fit to the
gas-phase abundances of Mg vrs Fe suggests that N(\FeII) $\propto$
N(\MgII)$^{0.57}$, which indicates that grain destruction returns more
Mg than Fe to the gas.  Although H ionization varies by up to 66\% in
the nearby ISM, Fe, Mg, and Si are mainly singly ionized (Section
\ref{sec:depletion}, \cite{SlavinFrisch:2008}), so that \FeII/\MgII\
ratio is relatively insensitive to ionization variations.  This
scenario suggests the presence of additional iron-rich grains in the
local ISM, with iron oxides a likely possibility
\citep[][]{Jones:1990feiron}.  Typical depletion patterns of Fe, Mg,
and Si shown in Figure \ref{fig:mgfe}, right, also suggest the
presence of at least two grain components, one consistent with
amorphous olivines (MgFeSiO$_4$) where $\mathrm{(Mg_{dust} +
Fe_{dust}) / Si_{dust} \approx 2}$, and an additional iron-rich
component in the dust.  A third characteristic of the depletion
pattern appears when depletions in tenuous and translucent clouds are
combined to display the gas-to-dust mass ratio, \Rgd, as a function of
the percentage of the dust mass carried by Fe (Figure
\ref{fig:cloudy}, left, and \cite{FrischSlavin:2003}).  The increase
of \Rgd\ with Fe-dominance of the core indicates that grain
destruction preferentially destroys lighter grain mantle material and
leave behind iron-rich cores (also see Section \ref{sec:js}).
Consistent with these characteristics are that gas-phase interstellar
Mg abundances are found to scale with Si abundances, and sometimes
independently of Fe abundances, both in low density galactic
sightlines and in the Small Magellenic Cloud, indicating that there is
a separate Fe-rich grain component
\citep{Howketal:1999,Welty:2001apjlsmc,Sofia:2006smc,Cartledge:2006abun}.
Frisch et al. \cite{Frischetal:1999} found that the composition of
local grains are consistent with olivine cores and pyroxene mantles.

Kimura and Mann \cite{KimuraMann:2003comp} looked in detail at the
local grain mineralogy, using the assumption that the local
interstellar cloud is 30\% ionized.  They found that the grain
mineralogy is quite sensitive to the assumed solar abundances.  By
comparing depletion patterns with likely grain models and the
condensation sequence of minerals, a viable model of a LIC grain
contains an Mg-rich olivine and pyroxene core with Fe-rich kamacite,
and mantles composed of organic refractory compounds of photoprocessed
C, H, O and N ices.  For this scenario, $\sim 45$\% of the oxygen is
depleted primarily into organic refractory mantles (CHON) and
enstatite (MgSiO$_3$), and the grain composition is found to be
consistent with grains in molecular clouds.

\subsection{Elemental depletion patterns for interstellar gas entering
heliosphere} \label{sec:depletion}

A unique aspect of the ISM enveloping the heliosphere is that the
cloud properties can be derived from \emph{in situ} data as well as
sightline-integrated observations of nearby stars.  Measurements of
interstellar \HeI\ and other interstellar neutrals inside of the
heliosphere sample the ionizations of the gas entering the
heliosphere, while measurements of the LIC towards nearby stars gives
a complete picture of sightline-integrated ionization.
Photoionization models of the LIC therefore reconstruct the full
gas-phase abundances of refractory elements, which are predominantly
singly ionized in the local ISM.

\begin{figure}[!b]
\begin{center}
    \makebox{\includegraphics[width=0.46\textwidth]{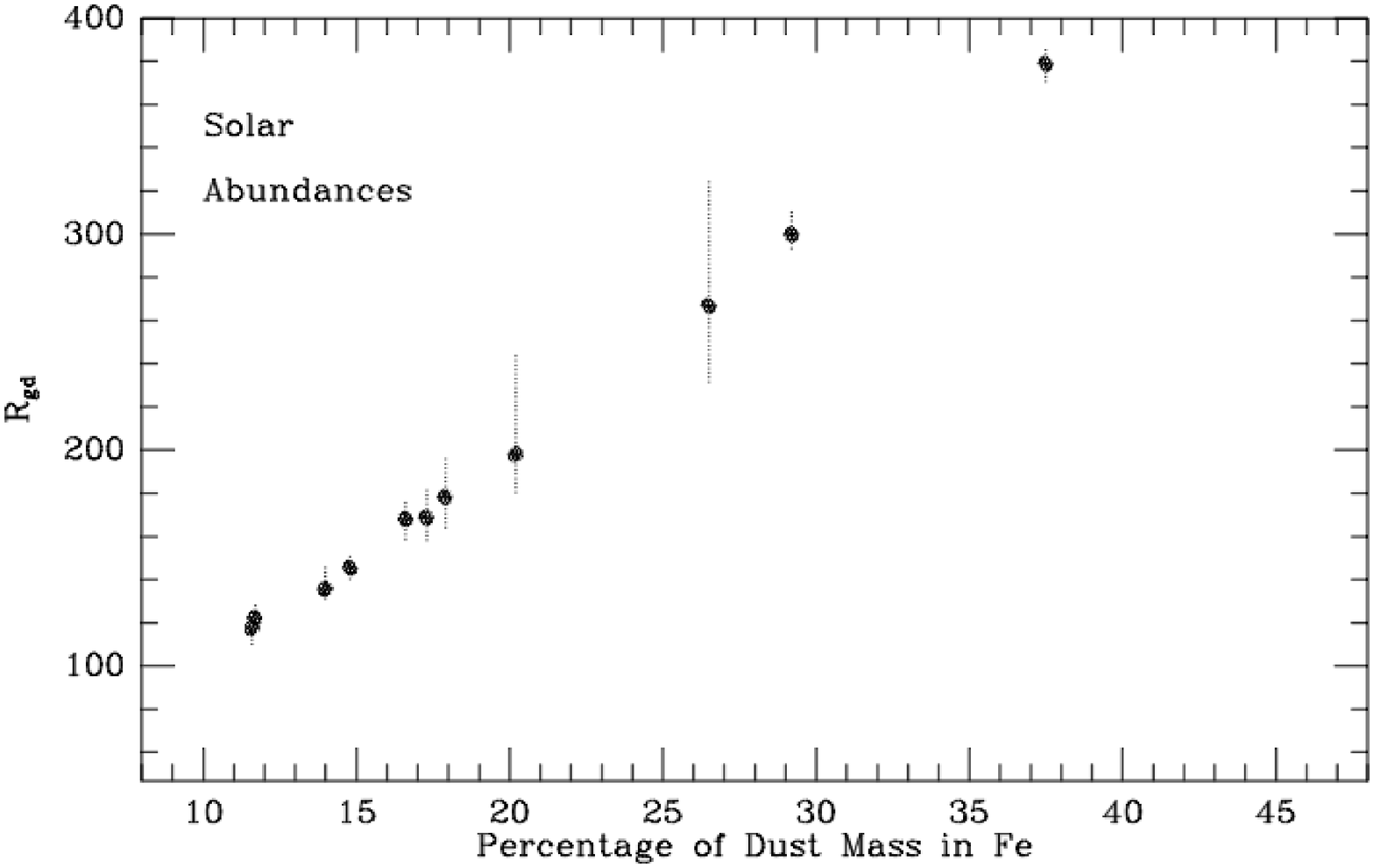}}
    \makebox{\includegraphics[width=0.46\textwidth]{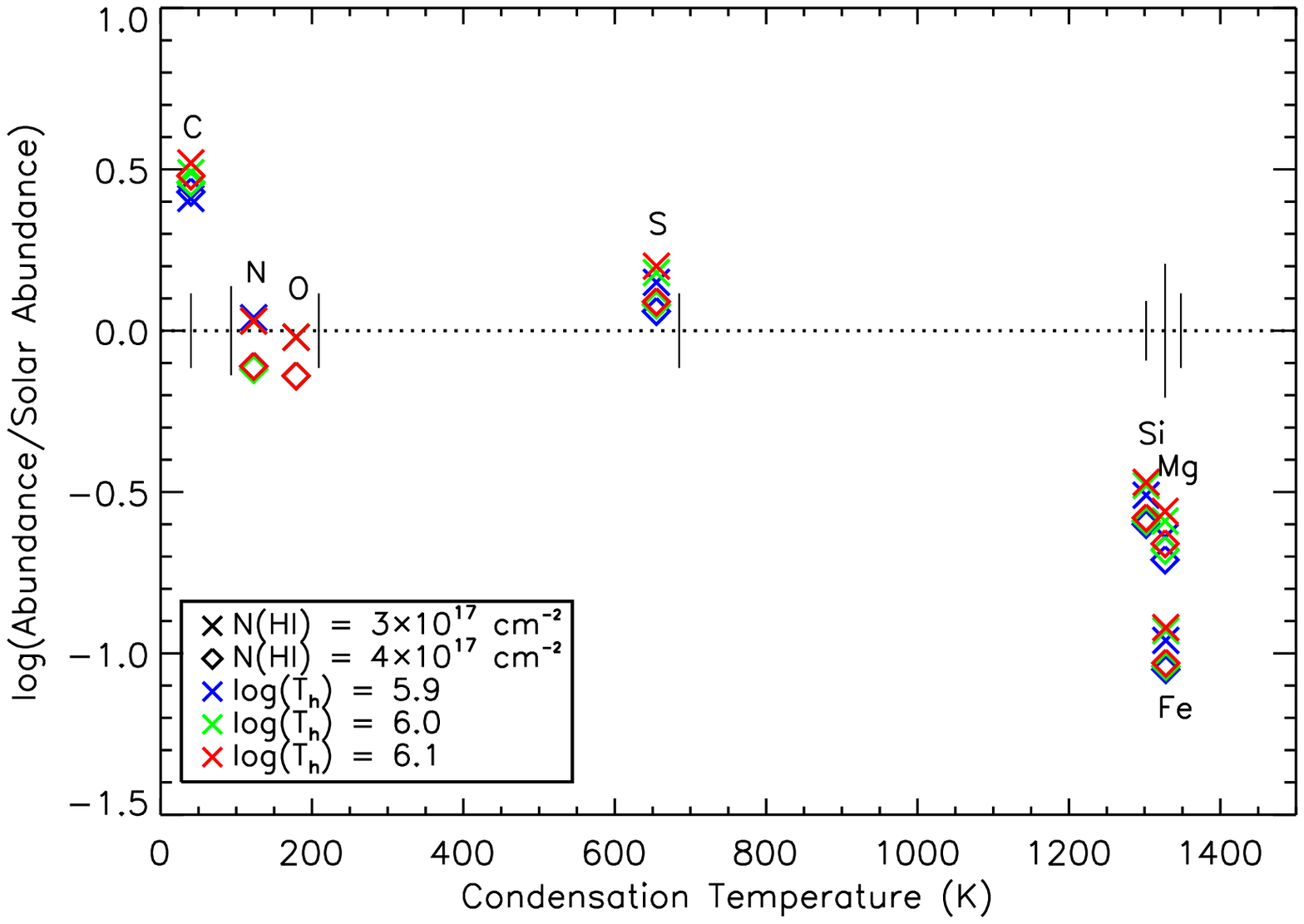}}
\end{center}
\vspace*{-5mm}
\caption[Depletions in the LIC interacting with the heliosphere:]{
Left: Percentage of dust mass carried by Fe as a function ol the
gas-to-dust mass ratio \Rgd\ (from \cite{FrischSlavin:2003}).  The
interstellar column densities for these stars span three orders of
magnitude, with the smallest values of \Rgd\ corresponding to the
lowest percentages of Fe in the grain mass.  Right: Abundances
relative to Asplund et al. \cite{Asplundetal:2005} solar abundances
vs.  condensation temperature. Results from models of the ionization
for the $\epsilon$ CMa line of sight including radiative transfer
(Model 26 in \cite{SlavinFrisch:2008}).  The symbol shape indicates
the assumed H\,\textsc{i} column density for the model, while the
color indicates the assumed temperature for the hot gas of the Local
Bubble.  The abundances are fixed so as to match the observed ion
column densities.  An overabundance of carbon in the gas phase is
required to match the observed carbon abundances, suggesting that
carbonaceous dust grains have been destroyed in the LIC.  The carbon
abundance in these models has been reconstructed from the cloud
ionization and the 1335.7 \AA\ carbon fine-struture line (see
\cite{SlavinFrisch:2008} for additional information).  }
\label{fig:cloudy}
\end{figure}

Model 26 in Slavin and Frisch \cite{SlavinFrisch:2008} provides a
viable model of the depletion patterns of the ISM seen inside and
outside of the heliosphere (Figure \ref{fig:cloudy}).  In this model
hydrogen is 22\% ionized, helium is 39\% ionized, and 15\% of the
magnesium and neon are doubly ionized.  These ionization levels show
that full ionization models are required to reconstruct the depletion
pattern for low density gas.  For this model, oxygen has a total (ions
plus neutrals) gas phase abundance of 331 ppm, so that 35\% of the
oxygen is depleted onto dust grains.  The He ionization is provided by
radiation generated in a conductive interface at the boundary of the
LIC and surrounding hot gas, as well as starlight from nearby white
dwarfs.  The depletion pattern in the LIC is consistent with grains
consisting of amorphous olivine.  These photoionization models yield a
supersolar carbon abundance in the LIC gas at the heliosphere, leaving
no carbon for the dust (Figure \ref{fig:cloudy}) and indicating that
the carbonaceous grains have been destroyed in the LIC (Section
\ref{sec:shock}).  Reconstruction of the gas-phase abundances in the
LIC gives a value for the gas-to-dust mass ratio, \Rgd, that
explicitly includes hidden ionized gas.  These models predict values
for \Rgd\ that are consistent with \emph{in situ} measurements (see
below).

\subsection{Dust composition from IBEX measurements of interstellar neon and
oxygen}

A new diagnostic of the oxygen composition of ISDGs has recently been
provided by \emph{in situ} measurements of interstellar \NeI\ and \OI\
by the Interstellar Boundary Explorer (IBEX) mission
\cite{Bochsleretal:2012isn}.  Interstellar \NeI\ and \OI\ flow into
the inner heliosphere from the upwind direction, near the galactic
center direction, where a special observing configuration allows the
neutral flow to be measured by the IBEX-LO instrument. From these data
an interstellar ratio of (Ne/O) $=0.27 \pm 0.10 $ is obtained. In
order to obtain this value, the measured particle fluxes are corrected
for propagation through the heliosphere, ionization of the neutrals in
the heliosphere and heliosheath regions (the latter termed
``filtration''), and the ionizations of Ne and O in the ISM
surrounding the heliosphere (which are provided by photoionization
models of the LIC \cite{SlavinFrisch:2008}).  The solar ratio is
(Ne/O) $\sim 0.17$.  Since Ne will be undepleted in the ISM, the IBEX
Ne/O ratio therefore indicates that $\sim 40$\% of the interstellar
oxygen in the ISM surrounding the heliosphere has been depleted onto
interstellar dust grains.

Both the photoionization models and \emph{in situ} Ne/O data indicate
that 35\%--40\% of the oxygen in the LIC is missing from the gas phase
and therefore depleted onto dust grains.  This pattern is consistent
with the global depletion pattern of oxygen in the ISM. Jenkins
\cite{Jenkins:2009dust} has found that oxygen depletes more rapidly
then refractories such as Mg, Si and Fe for all levels of depletions.
In the low density local ISM, it is difficult to identify the
compounds that lock up so many oxygen atoms in a solid phase since
water ices are not observed in high abundances.  Jenkins argues that
water ices may coat grains larger than 1 \micron\ that are difficult
to detect.  Pure metallic iron is unlikely in the ISM because of rapid
oxidation, but iron oxides are a possible hidden reservoir for oxygen
\cite{Jones:1990feiron}.  The missing O may be in organic refractory
grains such as seen by Stardust
\cite{Whittet:2010,KimuraMann:2003comp}.  Whittet et
al. \cite{Whittet:2010} argue that the oxidation ratios of metals in
dust would correspond to $\mathrm{O/(Mg+Si+Fe)} =1.5$ for a grain
population of $\mathrm{Mg Si O_3}$ silicates and $\mathrm{Fe_2 O_3}$
oxides and 1.2 for a mixture of $\mathrm{Fe O}$ and $\mathrm{Mg O}$
monoxides and primarily olivine silicates.  For the LIC, where the
level of ionization in the gas is reconstructed with photoionization
models, an oxidation ratio of 1.8 is found from model 26 in Slavin and
Frisch \cite{SlavinFrisch:2008} using abundances from Asplund et
al. \cite{Asplund:2009araa}.  Observations of six clouds towards four
nearby stars within 32 pc give a mean oxidation ratio of $2.5 \pm
0.7$.

The process of inferring the oxygen budget in the ISM is subject to
several major uncertainties, including the correct reference abundance
in the ISM and the composition of grains surviving shock destruction.
Based on the isotopic composition of the anomalous cosmic ray
component, the ISM flowing into the heliosphere has a solar
composition (Section \ref{sec:composition}). The solar oxygen
abundance is $490 \pm 58$ ppm, which is twice the oxygen abundance in
meteorites \cite{Asplund:2009araa}.  Cartledge et
al. \cite{Cartledge:2006abun} concluded that the problem with
``missing oxygen'' is minimized for an assumed reference standard for
the ISM of young F and G-star abundances, and a grain population of
silicates and $\mathrm{FeO}$ oxides.

\section{Interstellar Grains in the Heliosphere} \label{sec:insitu}

Interstellar dust grains have been measured within 5 AU of the Sun by
spacecraft \cite{Mann:2009araa}. Interstellar gas and dust flowing
into the heliosphere have the same velocity vectors when extrapolated
to the ISM at ``infinity''.  \emph{Ulysses} data give a dust velocity
at infinity of $24.5^{+1.1}_{-1.2}$ \kms\ \cite{KimuraMann:2003clic}
towards galactic coordinates $\ell,b=188^\circ \pm 15^\circ,-14^\circ
\pm 4^\circ$ \cite{Frischetal:1999}.  The gas velocity at infinity,
based on the concensus IBEX-LO observations of interstellar \HeI\ in
the inner heliosphere, is $23.2 \pm 0.3$ \kms\ towards the direction
$\ell,b=185.25^\circ \pm 0.24^\circ, -12.03^\circ \pm 0.51^\circ$,
with a temperature of $6,300 \pm 390$ K \cite{McComas:2012bow}.  A bow
shock around the heliosphere partially decouples the gas and dust
components of the ISM, so the dynamical interaction of small grains
with the heliosphere, in particular, is sensitive to the presence of a
bow shock.  For the new IBEX-LO LIC velocity vector, the heliosphere
will not have a bow shock for magnetic field strengths $> 3$ \microG.

In principle, if the propagation of interstellar dust grains through
the heliosphere is known, the grain mass distribution (for the larger
grains) in the ISM can be recovered.  A nominal power-law MRN
distribution \cite{MRN:1977} is assumed for this comparison.  Although
the MRN is not a realistic description of the extinction and emission
properties of interstellar dust over the full spectral range from the
infrared to ultraviolet, it plays a key role in the interpretation of
astronomical dust data \cite{ZubkoDwekArendt:2004}.  In the absence of
reliable data on the mass distribution of the dust in the ISM outside
of the heliosphere, the heliospheric modulation of interstellar dust
is often discussed in terms of the MRN distribution.  (We note that
the calculations of grain trajectories do not depend on the size
distribution.)  A distinct property of the mass distribution of
interstellar grains in the heliosphere is the deficit of small grains,
mass $< 10^{-15}$ gr (radius 0.04--0.2 \micron\ for density 2.5 gr
\cc), and an excess of large grains, masses $> 10^{-10}$ gr.  The
deficit is caused by the deflection of the small grains around the
heliosphere
\cite{Frischetal:1999,Slavinetal:2010tiny,Slavinetal:2011dustaas},
while the excess of large grains, radii $> 1$ \micron\ for a nominal
density of 2.5 \cc, signals an unexpected component of local
interstellar dust.  In 2005/2006 the dust flow direction shifted
southwards by $\sim 30^\circ$, probably due to the change in the solar
wind magnetic polarity \cite{Kruegeretal:2010}.

The gas-to-dust mass ratio, \Rgd, can be independently determined from
the \emph{in situ} dust measurements, and then compared with the
predictions of the LIC photoionization models.  A value of \Rgd$\sim
150$ is consistent both with recent \emph{Ulysses} data (H. Krueger,
private communication) and the photoionization models (which give a
range of \Rgd $=140-320$, depending on the selected solar abundance
\cite{SlavinFrisch:2008}).  The \emph{in situ} value for \Rgd\ only
depends on the photoionization models for the total gas mass, and
otherwise this \Rgd\ should be an upper limit because small grains are
excluded from the heliosphere.  If the values for \Rgd\ determined
separately from the dust measurements and observations of the gas
(combined with the missing-mass argument) do not agree, it would
indicate that the grains and gas are from different reservoirs, which
is to say a decoupling of gas and dust leading to a local dust
enhancement.  First indications are that \Rgd\ in the LIC as derived
from the \emph{in situ} dust measurements will decrease significantly
once the mass of the excluded grains is included.

Interstellar dust grains encountering the heliosphere will interact
most strongly with the magnetic field.  The trajectories resulting
from this interaction can be quite complex, depending on the
charge-to-mass ratio for the grain and the detailed morphology of the
magnetic field throughout the heliosphere.  The grains are generally
charged positively because of the strength of the solar UV/optical
radiation field which produces photoejection of electrons from the
grains.  We have used numerical magneto-hydrodynamical (MHD) $+$
kinetic (for the neutrals) calculations of the heliosphere to study
the trajectories of interstellar grains incident on the heliosphere
\citep{Slavinetal:2010tiny,Slavinetal:2011dustaas,Slavinetal:2012}.
We used two different solar wind magnetic field (SWMF) polarities in
our modeling, a de-focusing configuration (south pole positive,
corresponding to the current polarity) and focusing polarity (north
pole positive).

\begin{figure}[ht!]
\begin{center}
  	\makebox{\includegraphics[width=10cm]{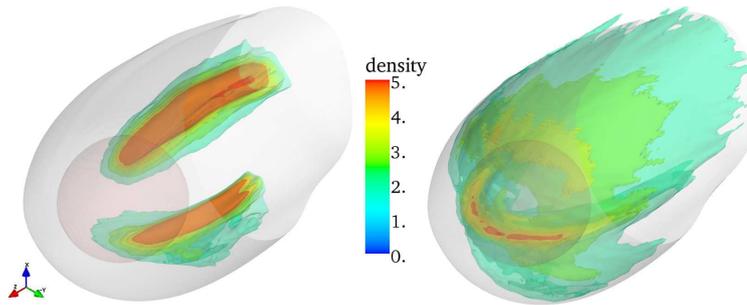}}
\end{center}
\caption{Dust density distribution for 0.18 \micron\ grains in the
heliosphere rendered as the 3-D surfaces corresponding to factors of
2, 2.75, 3.5, 4.25 and 5 enhancement over the ambient density.  The
de-focusing SWMF case is shown on the left and the focusing SWMF
polarity case is shown on the right.  The focusing SWMF leads to
enhanced density in a crescent shaped region near the ecliptic plane
while the de-focusing field leads to concentration of the dust at the
north and south ecliptic poles. The inner, nearly spherical surface is
the solar wind termination shock. Note: the interstellar inflow comes
from the lower left and flows toward the upper right.(Figure from
\cite{Slavinetal:2012})}
\label{fig:ddens_3D}
\end{figure}

Interstellar dust enters the the heliosphere region at the relative
speed of the solar system and the LIC. The smallest grains ($\sim 0.01
\mu$m) which have gyroradii that are small on the scale of the
heliosphere ($\lesssim 1$ AU) couple tightly to the interstellar field
and their trajectories follow that field as it is distorted by the
heliosphere.  Somewhat larger grains ($\sim 0.1 \mu$m) have gyroradii
large enough to allow them to penetrate the heliopause, at which point
their trajectories depend on the details, in particular the polarity,
of the SWMF.  If the field has magnetic north coinciding with the
Sun's north pole, the grains are deflected away from the ecliptic
while for the opposite polarity, they are focused in the ecliptic
plane.  For intermediate sized grains, $\sim 0.05 - 0.2\,\mu$m, the
SWMF polarity has strong effects on the grain trajectories and
resulting grain density, either concentrating the grains near the
north and south ecliptic poles (de-focusing polarity) or in the
ecliptic plane (focusing polarity).  Fig. \ref{fig:ddens_3D} shows a
3-D rendering of the dust density distribution for 0.18 $\mu$m grains.
For grains that are even larger, their gyroradii can be $\sim 1000$ AU
or larger and tend to be only slightly deflected from their initial
trajectory.  For these grains the effect of the Sun's gravity leads to
an enhanced space density in the inner solar system.

\begin{figure}[ht!]
\begin{center}
  \makebox{\includegraphics[width=10cm]{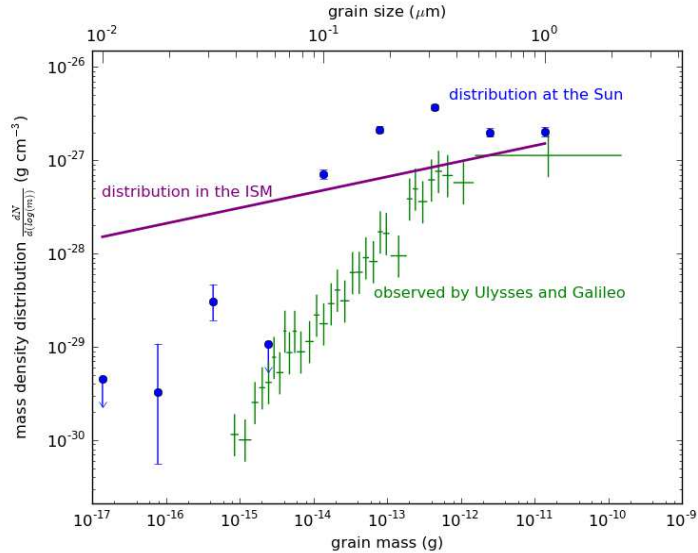}}
\end{center}
\caption{Grain mass distribution for interstellar grains in the
heliosphere observed by the \emph{Ulysses} and \emph{Galileo}
spacecrafts (updated, H.  Krueger private communication).  As an
example, the purple line shows the MRN power law size distribution
\citep{MRN:1977} but with higher upper and lower size cutoffs and
using a gas-to-dust mass ratio of 150.  The blue points show the grain
mass distribution that results in our model (focusing SWMF).  This
model clearly overpredicts the density of large grains, but is
promising for linking the observed grain size distribution with the
initial one in the Local Interstellar Cloud.}
\label{fig:size_distn}
\end{figure}

The very different density distributions of the grains in the models
that use different SWMF polarities indicate that we need to take into
account the time evolution of the heliosphere in such calculations.
An interstellar grain will take $\sim 30$ yr to go from the heliopause
to the Sun and so will experience two polarity changes because of the
solar cycle.  The SWMF polarity of the 1990's was de-focusing, while
that of the 2000's was focusing.  By calculating the enhancement in
grain space density at the Sun and comparing with the \emph{in situ}
dust detected with \emph{Ulysses} and \emph{Galileo}, we can, in
principle, infer the interstellar grain content, at least for the
larger grains.  In Fig.  \ref{fig:size_distn}, as an example, we show
the grain size distribution that an MRN \citep{MRN:1977} type size
distribution in the ISM would produce at the Sun under the assumption
of a focusing polarity in the SWMF.  Here the size distribution has
the MRN power law in size, $dn/da \propto a_\mathrm{grain}^{-3.5}$
(but we show the differential mass per logarithmic grain mass
distribution), though the grain size limits extend from 0.01 to 1.0
$\mu$m instead of the size limits used to fit interstellar extinction
which go from $\sim 0.005 \mu$m to $0.25 \mu$m.  A gas-to-dust mass
ratio of 150 is assumed to set the overall normalization of the
distribution.  It's clear from the figure that this ``shifted MRN''
distribution overpredicts the dust density at the Sun. As mentioned
above, however, the assumption of a single polarity for the SWMF over
the entire course of a grain trajectory is not realistic.  Future
modeling will need to take the time varying SWMF into account.

\section{Life cycle of local interstellar dust grains} \label{sec:shock}

Variations in the relative abundances of \MgII\ and \FeII\ in the gas
by a factor of 30 between different CLIC velocity components (Figure
\ref{fig:mgfe}) indicates an inhomogeneous grain population.  Either
multiple grain sources or the non-uniform destruction of the initial
grain population may account for these differences.  The kinematical
coupling between the gas and dust suggests that the local dust is far
removed from the original source in active stars and supernova, or
massive molecular clouds.  Rather, it appears that there have been
different levels of grain destruction in the CLIC, and that possibly
grain destruction is ongoing.  The very nature of the dispersion of
velocities in the CLIC ($\pm 4.6$ \kms\ about the bulk flow vector,
\cite{Frisch:2011araa}, Figure \ref{fig:localfluff}) suggests that
local dust grains may not have been uniformly shocked. In the Fe
core-silicate mantle model, shock processing that removes only the
mantle, such as shocks with velocities below 100 km s$^{-1}$, will
preferentially return C, Si and Mg to the gas phase as compared to Fe
by thermal sputtering (Figure \ref{fig:shock_dest_core-mantle}).  Iron
oxides may survive shocks in the cores of grains that have protective
mantles of a lower density material.

The efficiency of grain destruction by shocks depends on the
properties of the medium into which it propagates. If the pre-shock
gas were inhomogeneous, the efficiencies of grain destruction will
have varied.  For instance, increasing the cloud density by a factor
of ten raises the grain destruction rate by a factor of $\sim 1.5$,
while increasing the interstellar magnetic field strength from 1
\microG\ to 3 \microG\ reduces the grain destruction rate by factor of
$\sim 2$.

\subsection{Dust Destruction in the CLIC} \label{sec:js}

Our photoionization models for the LIC \cite{SlavinFrisch:2008} use
the observed LIC column densities toward $\epsilon$ CMa to set the gas
phase elemental abundances.  We then compare the abundances derived
this way with a set of solar abundances to indicate the amount of
depletion into grains.  As can be seen in Fig. \ref{fig:cloudy}, we
find modest depletions of the refractory elements that make up
silicate grains (Fe, Mg, Si, and O) and no depletion of C.  This low
level of depletion in the LIC can be interpreted as evidence of
partial destruction of grains by a fast shock sometime in the past.

\begin{figure}[ht!]
\begin{center}
 \makebox{\includegraphics[width=6.5cm]{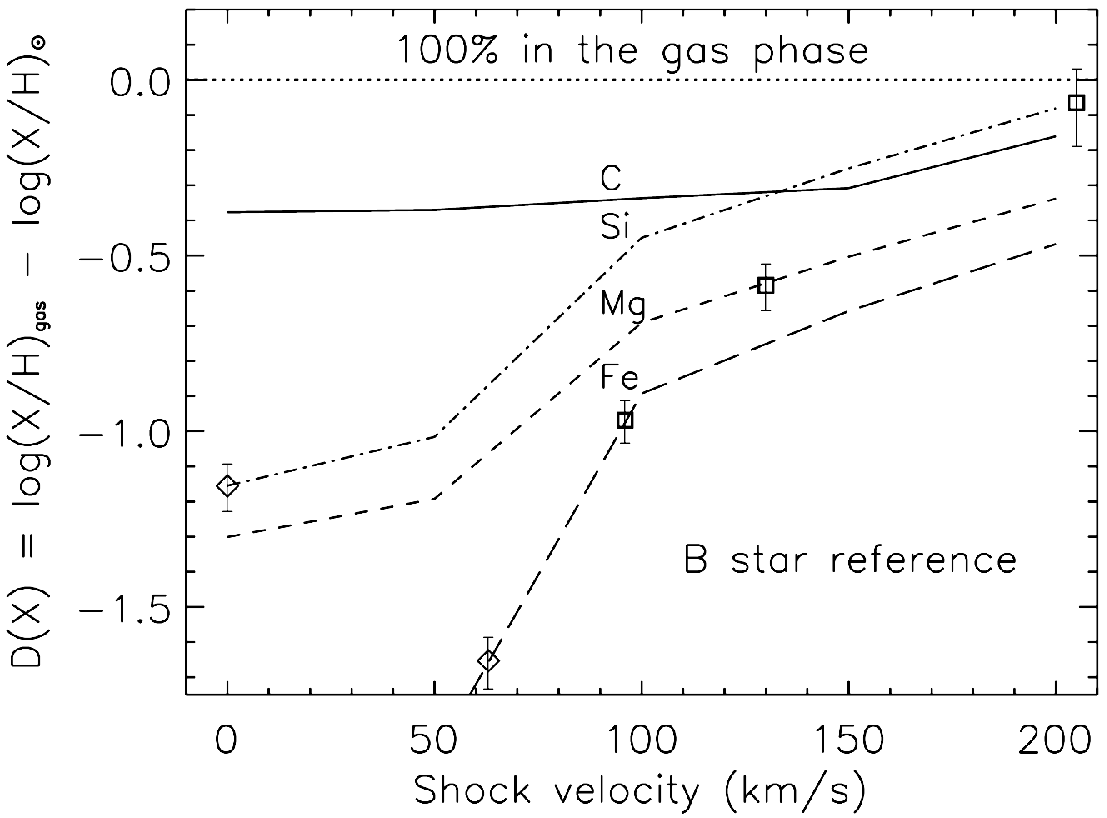}}
 \makebox{\includegraphics[width=7cm]{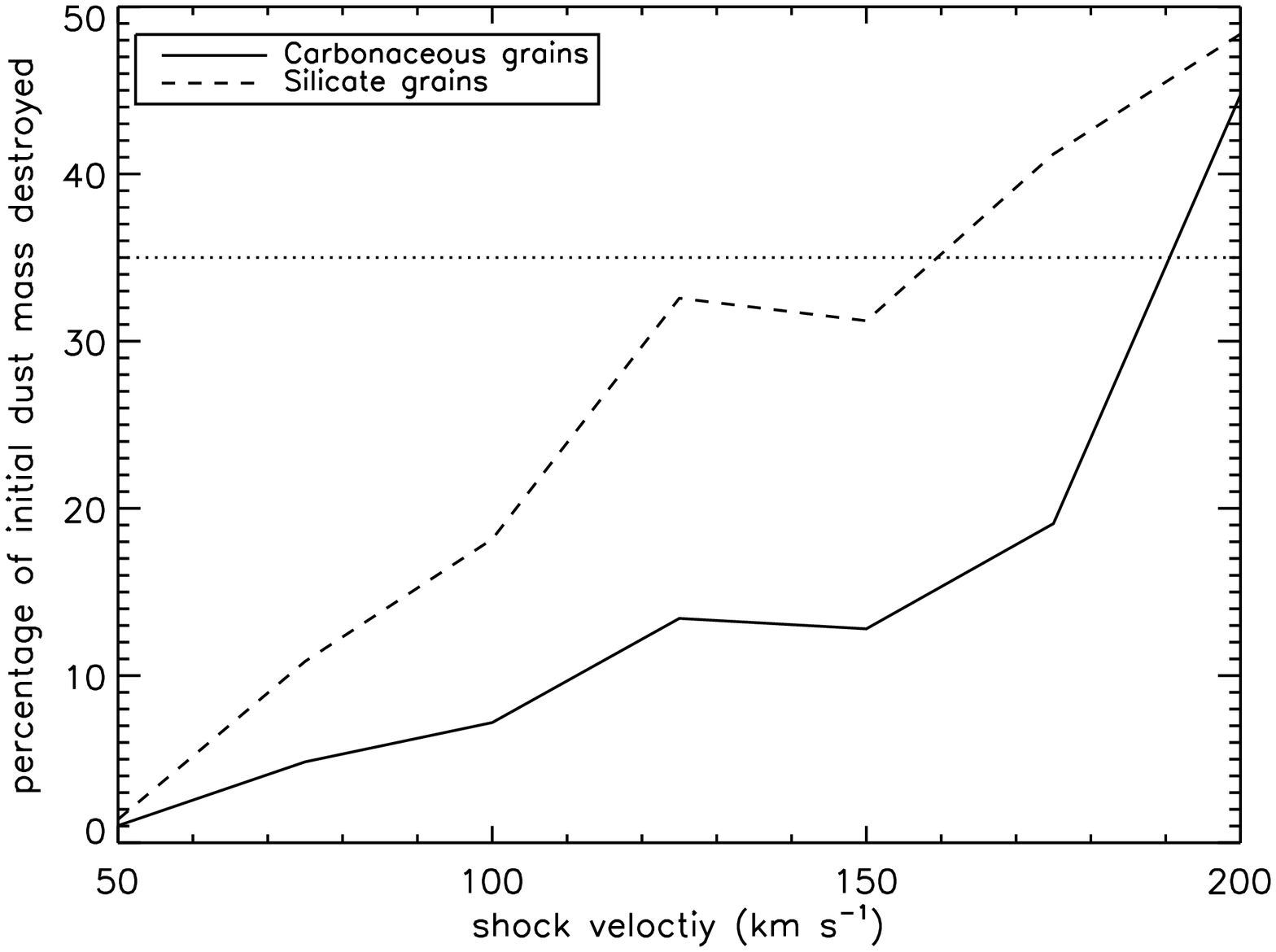}}

\end{center}
\caption{Left: Post-shock depletion vs.\ shock velocity for
core-mantle grain model where core is assumed to be silicate (olivine)
and mantle is pyroxene (from \cite{Frischetal:1999}).  Right:
Percentage of initial total grain mass lost because of dust
destruction in the shock vs.\ shock speed.  To get the Si gas phase
abundance determined for the LIC we need $\lesssim$35\% destruction of
silicate dust which is consistent with $v_\mathrm{shock} \lesssim 150$
km s$^{-1}$.  Such shocks should destroy less than 15\% of the
carbonaceous dust, however, in contradiction with the large gas phase
abundance of carbon found in many studies (also see Section
\ref{sec:depletion}).}
\label{fig:shock_dest_core-mantle}
\end{figure}

In relatively slow shocks, $v_\mathrm{shock} \sim 50 -100$ km
s$^{-1}$, the grain destruction is primarily via grain-grain
collisions caused by the betatron acceleration of the grains in the
post-shock cooling region.  This sort of destruction tends to erode
any mantle the grains might have and shift the grain size distribution
towards smaller grains.  In faster shocks, thermal and non-thermal
sputtering preferentially destroy smaller grains by removing the same
thickness of grain material regardless of grain size.  In Figure
\ref{fig:shock_dest_core-mantle}, left, we show the resulting
depletions after shock passage for a core-mantle grain model wherein
the core is silicate and mantle is pyroxene (from
\cite{Frischetal:1999}).  The effects of shock passage depend on, in
addition to assumptions about grain composition and structure, the gas
density and magnetic field strength.

The abundances derived for the LIC present a challenge for these dust
destruction models since the prediction is that carbonaceous dust will
be less destroyed than silicate dust for shock velocities over 150
\kms.  The depletions we find for silicate grain elements imply that
$\sim$35\% of the grain mass has been destroyed as could occur if a
shock of $\sim 150$ km s$^{-1}$ has passed through the cloud, but as
can be seen from Fig.  \ref{fig:shock_dest_core-mantle}, right, the
shock destruction models indicate that $<$15\% of the carbonaceous
grains should have been destroyed by such a shock.  A similar pattern
of high carbon dust destruction and modest silicate dust destruction
is inferred for lines of sight toward high and intermediate velocity
clouds \cite{Welty:2002zeta}.  These results prompted Jones and Nuth
\cite{JonesNuth:2011} to propose that instead of graphite as the model
for carbonaceous grains, hydrogenated amorphous carbon might be closer
to their structure.  Calculations they did for these grains indicate
that they are much easier to destroy in shocks, consistent with the
above results.

\subsection{Source of very large interstellar grains}

Large and small interstellar dust grains in the LIC may originate in
different regions of the CLIC.  The gyroradius for 1 \micron\ silicate
grain is $\sim 1$ pc for a grain with a density of 3.3 gr \cc, charge
of 200, and traveling 23 \kms\ through a 3 \microG\ magnetic field.
Small, radius $a \sim 0.1$ \micron, grains originate locally. Grains
that are 1 \micron\ and larger may originate in more distant regions
of the CLIC.  The presence of large ISDGs in the ISM, radius $a > 1$
\micron, if widespread, would violate abundance constraints set by
interstellar extinction curves.  This suggests that they trace a
different ISM reservoir than the small grains \cite{Draine:2009}.  The
large gyroradius of the large grains allows for them to be decoupled
from the small grains so that the grain population may not be
uniformly shocked.

The large grains (radius $a > 1$ \micron) found by \emph{in situ}
measurements of interstellar dust suggests that molecular cloud grains
are mixed into the local ISM.  Infrared observations of dense
molecular clouds require micron-sized ISDGs to reproduce the
``coreshine'' emission.  Coreshine is produced when dusty clouds
scatter background 8 \micron\ radiation that is spatially coincident
with the cloud cores \cite{Pagani:2010dust}.  The only known dense
dust cloud near the Sun is a thin filamentary-like cold (30 K) feature
towards Leo.  The distance of this cloud has been constrained by
absorption lines in nearby stars to be within 11--24 pc
\cite{Peeketal:2011,Frisch:2011araa}.  Densities in the Leo cloud are
\NHI$\sim 10^{19}-10^{20}$ \cmtwo, and $n \sim 320 $ \cc.  The Leo
cloud velocity is consistent with the CLIC bulk velocity, so either
this cloud or other similar tiny dense local clouds
(e.g. \cite{Haud:2010}) may explain the large dust grains observed
inside of the heliosphere.

\section{Summary}

Local interstellar dust shows the characteristics of dust in low
density clouds, with high gas phase abundances of refractory elements
that have been processed by $\sim 150$ \kms\ shocks.  Grains consist
of silicates, primarily olivine, and an additional Fe-rich grain
component, possibly of complex oxides.  Observations of interstellar
oxygen both inside and outside of the heliosphere suggest high
depletions that indicate an additional carrier of the oxygen such as
either organic refractory particles or oxides.  The proximity of the
Sun to the Loop I superbubble suggests that grain destruction may have
occurred in shocks associated with that superbubble.
 
Large $> 1$ \micron\ interstellar dust grains have been detected in
the heliosphere, indicating either that some grains escaped
destruction in interstellar shocks, or that this dust has decoupled
from a nearby reservoir of large grains, or both.  The observed
deficit of small grains, $a < 0.1$ \micron, in the \emph{in situ}
grain sample indicates that the smallest grains are excluded from
heliospheric plasma because of large charge-to-mass ratios.  Models of
this exclusion indicates that plumes of $\sim 0.18$ \micron\ dust flow
around the heliosphere, with the exact configuration of the plumes
dependent on the solar wind magnetic field polarity and therefore on
the solar cycle phase.  These diverse techniques for studying local
interstellar dust provide an intriguing and still incomplete picture
of the properties of dust in low density regions of the galaxy.


\acknowledgments{This work has been supported by the Interstellar
Boundary Explorer mission as a part of NASA's Explorer Program,
and by NASA grant NNX08AJ33G to the University of Chicago.
The authors would like to thank an anonymous referee for emphasizing that
iron oxides help to explain the observed characteristics of the local
depletion patterns.}


\email{P.Frisch (e-mail: frisch@oddjob.uchicago.edu)}
\label{finalpage}
\lastpagesettings

\end{document}